\documentclass[graybox]{svmult}

\usepackage{type1cm}

\usepackage{makeidx}         
\usepackage{graphicx}        
\usepackage{multicol}        
\usepackage[bottom]{footmisc}

\usepackage{newtxtext}       %
\usepackage{newtxmath}       

\usepackage[numbers,sort&compress]{natbib}

\usepackage{amssymb}
\usepackage{amsmath}
\usepackage{upgreek}

\usepackage{hyperref}

\usepackage{graphicx}
\usepackage{pgfplots}
\usepackage{pgfplotstable}
\usepackage{bm}
\usepackage{url}
\usepackage{tikz}

\usepackage{subcaption}
\usepackage{overpic}
\usepackage{multirow}
\usepackage{lscape}
\usepackage{booktabs}
\usepackage{enumitem}
\usepackage[normalem]{ulem}

\usepackage{listings}

\lstdefinelanguage{myc++} 
{
  language=C++,
  captionpos=b,
  frame=single,
  keywordstyle=\bfseries\ttfamily,
  basicstyle=\scriptsize\ttfamily,
  commentstyle=\color{gray}\ttfamily,
  stringstyle=\rmfamily,
  numbers=left,
  numberstyle=\scriptsize,
  stepnumber=1,
  numbersep=8pt,
  breaklines=true,
  lineskip=1pt,
  frame=L,
  escapechar=\$,
  morekeywords = {constexpr},
}
\lstdefinelanguage{introspection} 
{
  captionpos=b,
  frame=single,
  keywordstyle=\bfseries\ttfamily,
  basicstyle=\scriptsize\ttfamily,
  commentstyle=\color{gray}\ttfamily,
  stringstyle=\rmfamily,
  numbers=left,
  numberstyle=\scriptsize,
  stepnumber=1,
  numbersep=8pt,
  breaklines=true,
  lineskip=1pt,
  frame=L,
  escapechar=\$
}
\newcommand*{\Comment}[1]{\hfill\makebox[7.0cm][l]{// \emph{#1}}}

\lstdefinelanguage{report} 
{
  captionpos=b,
  frame=single,
  keywordstyle=\bfseries\ttfamily\tiny,
  basicstyle=\scriptsize\ttfamily\tiny,
  commentstyle=\color{gray}\ttfamily,
  stringstyle=\rmfamily,
  numbers=left,
  numberstyle=\scriptsize,
  stepnumber=1,
  numbersep=8pt,
  breaklines=true,
  lineskip=1pt,
  frame=L,
  escapechar=\$
}

\usepackage{float}
\newfloat{listing}{tbp}{lop}  
\floatname{listing}{Listing}

\makeindex

\begin{document}

\title*{Large-Scale Simulations of Turbulent Flows using Lattice Boltzmann Methods on Heterogeneous High Performance Computers}
\titlerunning{Large-Scale Simulations of Turbulent Flows using Lattice Boltzmann Methods}
\author{Adrian Kummerländer, Fedor Bukreev, Yuji Shimojima, Shota Ito\\and Mathias J. Krause}
\authorrunning{Kummerländer, Bukreev, Shimojima, Ito, Krause}
\institute{Adrian Kummerländer \at Lattice Boltzmann Research Group (LBRG), Institute of Applied and Numerical Mathematics (IANM), Karlsruhe Institute of Technology (KIT) \email{kummerlaender@kit.edu}}

\maketitle

\abstract{Current GPU-accelerated supercomputers promise to enable large-scale simulations of turbulent flows. Lattice Boltzmann Methods (LBM) are particularly well-suited to fulfilling this promise due to their intrinsic compatibility with highly parallel execution on both SIMD CPUs and GPUs. A novel LBM scheme for wall-modeled LES in complex geometries is described with a special focus on the efficient implementation in the open source LBM framework OpenLB. Detailed scalability results are provided for all HoreKa partitions, utilizing up to 128 nodes and covering problem sizes up to 18 billion cells.}

\section{Introduction}

Multi-physics transport problems are predominant in the physical reality; capturing them in high-fidelity simulations enjoys ever-growing demand in both fundamental research and industrial applications.
This present report summarizes our efforts on modeling and simulating such problems on the HoreKa supercomputer using the open source software framework OpenLB~\cite{openlb2020,kummerlander_openlb_2025}. A particular focus is placed on the efficient implementation of large-scale simulations  of turbulent fluid flows on GPU clusters.

\begin{figure}[h]
\begin{subfigure}{0.49\textwidth}
    \includegraphics[width=\textwidth]{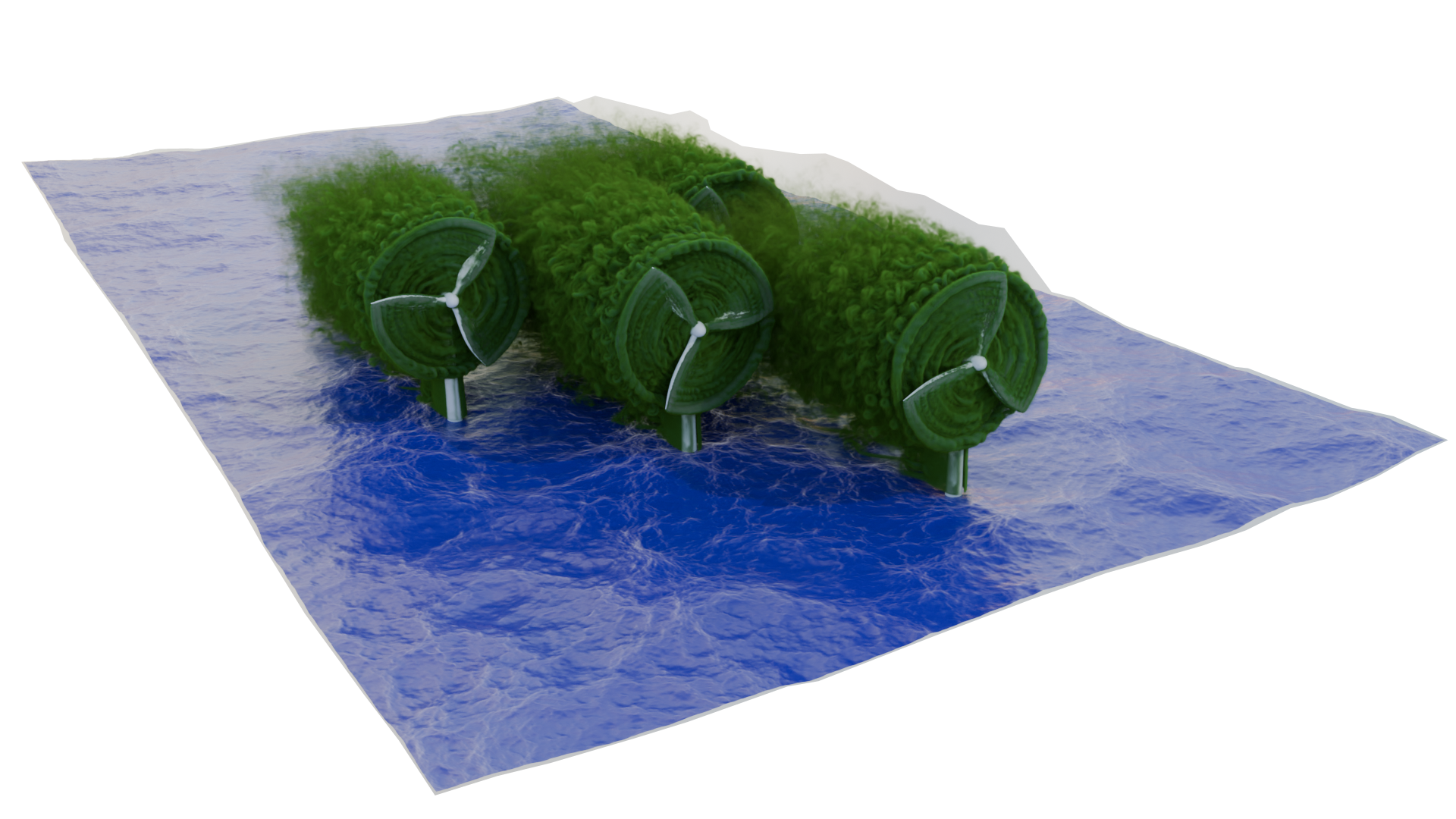}
    \caption{Fully-coupled FSI simulation of a wind park}
\end{subfigure}
\begin{subfigure}{0.49\textwidth}
    \includegraphics[width=\textwidth]{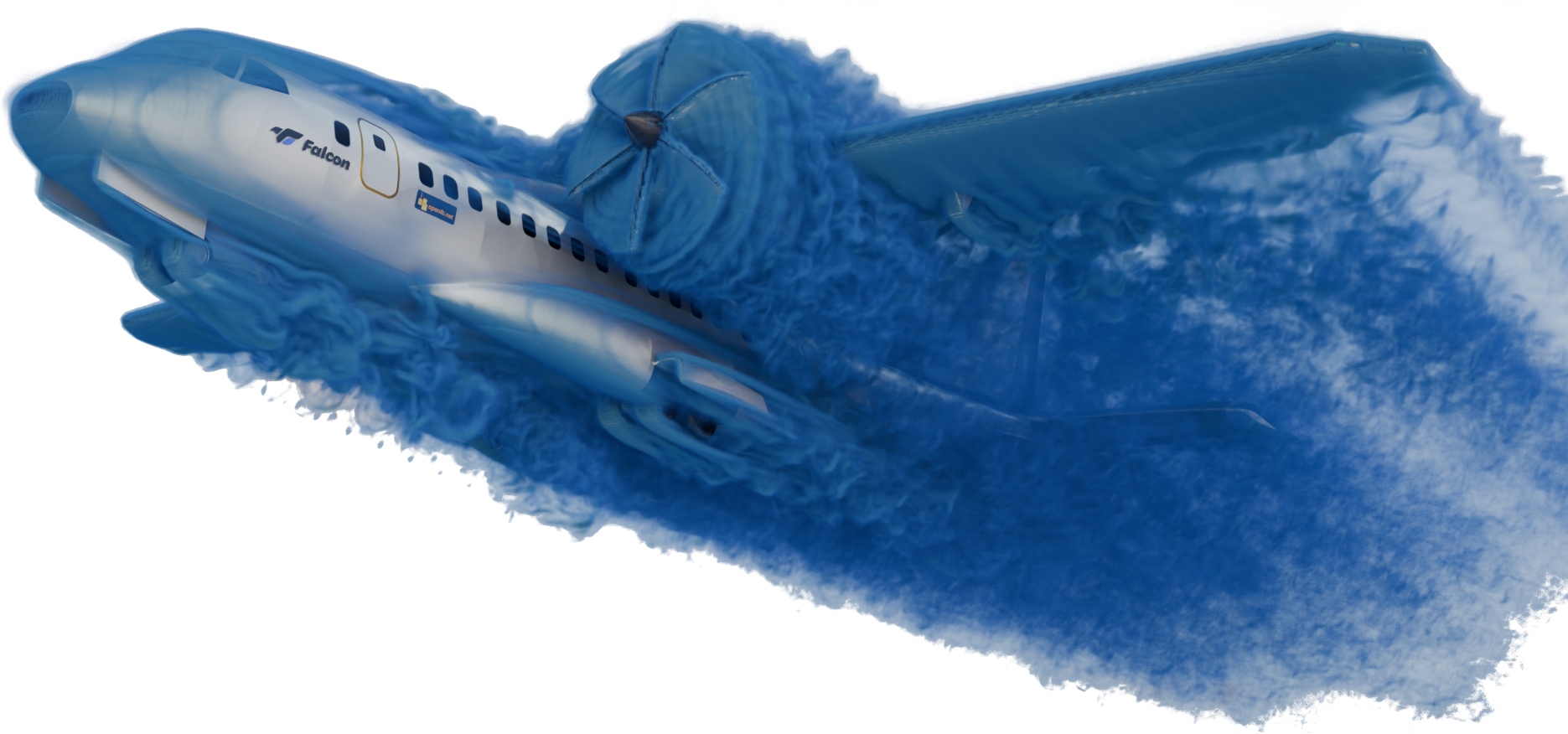}
    \caption{Wall-modelled FSI of a propeller aircraft}
\end{subfigure}
\caption{Volumetric rendering for large-scale multi-GPU LES in OpenLB}
\end{figure}

\emph{Reynolds‐averaged Navier–Stokes} (RANS) models and \emph{Large‐Eddy Simulation} (LES) are two often compared approaches to turbulent flow modeling: although RANS approaches outperform LES in terms of computational cost, they suffer from significant accuracy shortcomings in aerodynamic calculations~\cite{blocken_application_2011}.
Despite the availability of modern GPU-accelerated clusters such as HoreKa, LES is often considered as too expensive to replace RANS in practice.
Addressing the high computational cost of accurate LES for turbulent flows in large-scale, complex geometries is an ongoing research question.

Modern approaches such as \emph{lattice Boltzmann methods} (LBM)~\cite{chen_lattice_1998} offer a promising alternative to conventional \emph{finite‐volume methods} (FVM) for conducting LES.
OpenLB provides a platform-independent and fully differentiable framework~\cite{openlb2020,kummerlanderImplicitPropagationDirectly2023,Kummerlaender.deRSE23,kummerlaender25} for implementing and efficiently executing LBM-based simulations on heterogeneous high-performance computers.
In a fair comparison with the FVM solver OpenFOAM, it has previously demonstrated a 32× faster time-to-solution—even without leveraging its multi-GPU capabilities~\cite{HAUSSMANN20}.

As covering all recent applications~\cite{teutscher2025,Ito2025,Marquardt2025,Hettel2025,Zhong2024,SIMONIS23,Bukreev2025,Simonis2024,Marquardt2024a,Marquardt2024b,Ito2024,Prinz2024,Hafen2023,Bukreev2023}
of such a multi-physics simulation tool within LBRG, an interdisciplinary group of mathematicians and engineers is beyond the scope of a single report, we focus instead in detail on recent work specifically on modeling large-scale turbulent flows in urban areas~\cite{teutscher2025}. 
Studies~\cite{PASQUIER23,ahmad_large-eddy_2017,jacob2021,inagaki_numerical_2017,onodera_real-time_2021,lenz_towards_2019} show that accurate LES are fundamentally feasible even in larger built‐up areas. LBMs have already been used for evaluating wind comfort in complex urban geometries~\cite{ahmad_large-eddy_2017}, for simulating moist air convection or pressure loads on high‐rise buildings~\cite{jacob2021}, for analyzing atmospheric boundary‐layer dynamics~\cite{inagaki_numerical_2017}, or for investigating pollutant dispersion~\cite{onodera_real-time_2021}.
In particular, real‐time numerical simulations of urban flows with grid sizes on the order of one meter are possible~\cite{onodera_real-time_2021,lenz_towards_2019}.

\medskip
Section~\ref{sec:method} describes macroscopic equations used to model air flows of urban geometries, their solution by a novel LBM scheme and its efficient platform-transparent implementation in OpenLB.
Detailed parallel performance results on all partitions of HoreKa as well as a summary of the experimental validation of the developed numerical wind channel setup is provided in Section~\ref{sec:evaluation}. 

\section{Methodology}\label{sec:method}

Modeling air flows and pollution transport in urban geometries is particularly challenging for \emph{computational fluid dynamics} (CFD) due to the combination of large (landscape, buildings) and small length scales (building surfaces, trees).
In our digital twin study~\cite{teutscher2025} we modeled this complex problem using a filtered Brinkman--Navier--Stokes equations (FBNSE), capturing both LES for subgrid-scale turbulence modeling and homogenized porous media for tree modeling in a single target equation.
This equation was solved efficiently on HoreKa Green using a \emph{homogenized lattice Boltzmann method} (HLBM) discretization in OpenLB.

\subsection{Filtered Brinkmann--Navier--Stokes Equations}

The macroscopic motion of fluids is commonly described using the Navier--Stokes equations (NSE).
Incompressible flows in heterogeneous domains consisting of both fully fluid regions and porous media can be described using the filtered Brinkman--Navier--Stokes equations (FBNSE)
\begin{align}
\begin{cases}
    \bm{\nabla} \cdot \bar{\bm{u}}  =0, & \quad \text{in } \Omega \times I, \\
\frac{\partial \bar{\bm{u}}}{\partial t} + \bar{\bm{u}} \cdot \bm{\nabla} \bar{\bm{u}} = -\frac{\bm{\nabla} \bar{p}}{\rho} + \nu_{\mathrm{mo}} \bm{\nabla}^2 \bar{\bm{u}} + \frac{\nu_{\mathrm{mo}}}{K} \bar{\bm{u}} - \bm{\nabla} \cdot \mathbf{T}_{\mathrm{sgs}}, & \quad \text{in } \Omega \times I, \label{eq:fbnse}
\end{cases}
\end{align}
for filtered pressure \(\bar{p}\), velocity \(\bar{\bm{u}}\) density \(\rho\) 
on spatial domain \(\Omega \subseteq \mathbb{R}^3\) and time \(I\subseteq \mathbb{R}_{>0}\).
The molecular kinematic viscosity is defined as \(\nu_{\mathrm{mo}}\) and the permeability coefficient $K>0$ of the porous medium is given by the Forchheimer equation
\begin{equation}
K = \frac{\mu_F Q}{A \left(\frac{\Delta P}{L} - \frac{\rho}{K_{\beta}} \frac{Q^2}{A^2} \right)},
\end{equation}
with dynamic viscosity $\mu$, volume flow rate $Q$, characteristic length $L$, projected area $A$, pressure difference $\triangle P$ and nonlinear permeability coefficient $K_\beta$.
The term $\bm{\nabla} \cdot \mathbf{T}_{\mathrm{sgs}}$ models the subgrid-scale turbulence using the Smagorinsky LES approach  
\begin{align}
    \mathbf{T}_{\mathrm{sgs}} & = 2 \nu_\mathrm{turb} \bar{\mathbf{S}}, \label{eq:sgsStress}\\
    \nu_\mathrm{turb} & = \left(C_{\mathrm{S}} \triangle_{\bm{x}}\right)^2 \left|\bar{\mathbf{S}}\right|, \label{eq:turbVisc}
\end{align}  
where $C_{\mathrm{S}}>0$ is the Smagorinsky constant, $\triangle_{\bm{x}}$ is the filter width, and $\bar{\mathbf{S}}$ is the filtered strain rate tensor:  
\begin{equation}
    \bar{S}_{\alpha\beta} = \frac{1}{2} \left( \frac{\partial \bar{u}_{\alpha}}{\partial x_\beta} + \frac{\partial \bar{u}_{\beta}}{\partial x_\alpha} \right).
\end{equation} 

\subsection{Homogenized Lattice Boltzmann Method}\label{sec:hhrrlbm}

\begin{figure}
\centering
\begin{tikzpicture}[>=latex]
 \centering
    \draw[densely dashed] (-2,0,0) -- (0,0,0);
    \draw[densely dashed] (-2,0,0) -- (-2,2,0);
    \draw[densely dashed] (-2,0,0) -- (-2,0,2);
    \draw[densely dashed] (-2,0,2) -- (0,0,2);
    \draw[densely dashed] (-2,2,0) -- (0,2,0);
    \draw[densely dashed] (-2,0,2) -- (-2,2,2);
    \draw[densely dashed] (-2,2,0) -- (-2,2,2);
    \draw[densely dashed] (-2,2,2) -- (0,2,2);
    \draw[densely dashed] (0,0,0) -- (0,0,2);
    \draw[densely dashed] (0,0,0) -- (0,2,0);
    \draw[densely dashed] (0,0,2) -- (0,2,2);
    \draw[densely dashed] (0,2,2) -- (0,2,0);
 	\draw[densely dashed] (-2,0,1) -- (0,0,1); 
	\draw[densely dashed] (-2,0,1) -- (-2,2,1); 
	\draw[densely dashed] (-2,2,1) -- (0,2,1); 
	\draw[densely dashed] (0,2,1) -- (0,0,1); 
	\draw[densely dashed] (-1,0,0) -- (-1,0,2); 
	\draw[densely dashed] (-1,0,2) -- (-1,2,2); 
	\draw[densely dashed] (-1,2,2) -- (-1,2,0); 
	\draw[densely dashed] (-1,2,0) -- (-1,0,0); 
	\draw[densely dashed] (-2,1,0) -- (0,1,0); 
	\draw[densely dashed] (0,1,0) -- (0,1,2); 
	\draw[densely dashed] (0,1,2) -- (-2,1,2); 
	\draw[densely dashed] (-2,1,2) -- (-2,1,0);   
    \draw[->,thick](-1,1,1) -- (-1,1,0);
    \draw[->,thick](-1,1,1) -- (-2,1,0);
    \draw[->,thick](-1,1,1) -- (-1,2,0);
    \draw[->,thick](-1,1,1) -- (-1,0,0);
    \draw[->,thick](-1,1,1) -- (0,1,0);
    \draw[thick,fill=cyan](-1,1,0) circle(2pt);     
	\draw[thick,fill=green](-2,1,0) circle(1.5pt);
  	\draw[thick,fill=green](-1,2,0) circle(1.5pt);
  	\draw[thick,fill=green](-1,0,0) circle(1.5pt);
	\draw[thick,fill=green](0,1,0) circle(1.5pt);
    \draw[->,thick](-1,1,1) -- (-1,2,1);
    \draw[->,thick](-1,1,1) -- (0,1,1);
    \draw[->,thick](-1,1,1) -- (-2,1,1);
    \draw[->,thick](-1,1,1) -- (-1,0,1);
    \draw[->,thick](-1,1,1) -- (-2,0,1);
    \draw[->,thick](-1,1,1) -- (0,0,1);
    \draw[->,thick](-1,1,1) -- (-2,2,1);
    \draw[->,thick](-1,1,1) -- (0,2,1);
	\draw[thick,fill=cyan](-1,2,1) circle(2pt);
    \draw[thick,fill=cyan](0,1,1) circle(2pt);
    \draw[thick,fill=cyan](-1,0,1) circle(2pt);
    \draw[thick,fill=cyan](-2,1,1) circle(2pt);
    \draw[thick,fill=green](-2,0,1) circle(1.5pt);     
  	\draw[thick,fill=green](0,0,1) circle(1.5pt);
    \draw[thick,fill=green](-2,2,1) circle(1.5pt);     
  	\draw[thick,fill=green](0,2,1) circle(1.5pt);
	\draw[->,thick](-1,1,1) -- (-1,1,2);
    \draw[->,thick](-1,1,1) -- (-1,0,2);
	\draw[->,thick](-1,1,1) -- (-2,1,2);
	\draw[->,thick](-1,1,1) -- (0,1,2);
	\draw[->,thick](-1,1,1) -- (-1,2,2);
    \draw[thick,fill=cyan](-1,1,2) circle(2pt);
	\draw[thick,fill=green](-1,0,2) circle(1.5pt);
	\draw[thick,fill=green](-2,1,2) circle(1.5pt);
	\draw[thick,fill=green](0,1,2) circle(1.5pt);
	\draw[thick,fill=green](-1,2,2) circle(1.5pt);
    \draw[thick,fill=orange](-1,1,1) circle(3pt);
  \end{tikzpicture}
\hspace{2em}
\begin{tikzpicture}
	\draw[->,thick] (0,0,0) -- (.5,0,0);
    \node[anchor=north] at (.5,0,0) {$x$};
    \draw[->,thick] (0,0,0) -- (0,.5,0);
    \node[anchor=east] at (0,.5,0) {$y$};
	\draw[->,thick] (0,0,0) -- (0,0,.5);
	\node[anchor=north west] at (0,0,.5) {$z$};
\end{tikzpicture}
\caption{Schematic illustration of the discrete velocity set \(D3Q19\). Figure from \cite{simonis2023pde}.}
\label{fig:lattice}
\end{figure}
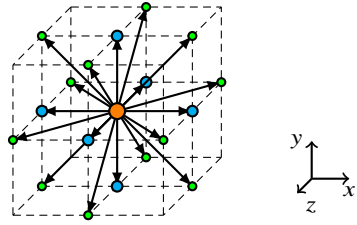

The HLBM is used to discretize the FBNSE~\eqref{eq:fbnse} on a regular space-time grid with the \(D3Q19\) velocity stencil (cf. Figure~\ref{fig:lattice}).
Specifically, we utilize a \emph{homogenized hybrid regularized recursive Lattice Boltzmann method with Smagorinsky LES model} (HHRRLBM-LES) that extends the classic HLBM~\cite{KRAUSE20171} with a hybrid third-order recursive regularized collision model~\cite{coreixas2017,JACOB18}.

The filtered and homogenized LB equation is given by  
\begin{equation}\label{eq:hlbm}
    f_{i} (\bm{x}+\xi_i \triangle t, t+\triangle t) 
    = 
    f_{i}^{\mathrm{eq}} (\bm{x}, t) + \left( 1 - \frac{1}{\tau_{\mathrm{eff}}(\bm{x},t)} \right)\tilde{f}_{i}^{(1)}(\bm{x}, t), \quad \text{in } \Omega_{\triangle x} \times I_{\triangle t}, 
\end{equation}
for distribution functions \(f_i\) along \(q\) discrete velocities \(\xi_i\) on a regular lattice \(\Omega_{\triangle x} \subset \Omega \subseteq \mathbb{R}^3\) with cell size \( \triangle x \) at discrete times \(I_{\triangle t} \subset I \subseteq \mathbb{R}_{\geq 0}\) separated by step size \(\triangle t\).
Here, the non-equilibrium distribution \(\tilde{f}_i^{(1)}\) is computed as a linear combination
\begin{align}
    \tilde{f}_{i}^{(1)}(\bm{x},t) = \sigma f_{i}^{(1)} (\bm{x},t)  - (1-\sigma) f_{i}^{(1,FD)} \text{ for } \sigma \in [0,1].
\end{align}
That is, the distribution is \emph{hybridized} between reconstructions using the rate of strain tensor obtained either from local macroscopic moments or from a non-local finite difference (FD) approximation.

For the local part, the non-equilibrium distribution function $f_{i}^{(1)}$ is expanded in terms of Hermite polynomials \(\mathbf{H}_i^{(n)}\) of the discrete velocity \(\xi_i\) as
\begin{equation}
  f_{i}^{(1)}(\bm{x},t) = \omega_i \sum_{n=0}^{N=3} \frac{1}{c_{\mathrm{s}}^{2n} n!} \mathbf{H}_i^{(n)} : \mathbf{a}_1^{(n)}(\bm{x},t) ,
\end{equation}
where \(\omega_i\) are the lattice weights. The Hermite expansion coefficients are defined as
\begin{equation}
    \mathbf{a}_1^{(n)}(\bm{x},t)=\sum_{i=0}^{q-1}\mathbf{H}_i^{(n)}f_i^{(1)}(\bm{x},t).
\end{equation}
For the non-local part, the FD non-equilibrium distribution function is defined as
\begin{equation}
f_{i}^{(1,FD)} := \frac{\rho \tau}{c_{\mathrm{s}}^{2}} \mathbf{H}_{i}^{(2)} : \mathbf{S}^{\mathrm{FD}} (\bm{x},t).
\end{equation}

The equilibrium distribution function is defined as
\begin{equation}
  f_i^{\mathrm{eq}}(\bm{x},t) = \omega_i \left(\rho + \frac{\xi_i \cdot \rho\, \mathbf{\widehat{u}}}{c_{\mathrm{s}}^2}
    + \frac{\mathbf{H}_i^{(2)} : \widehat{\mathbf{a}}_0^{(2)}}{2 c_{\mathrm{s}}^4}
    + \frac{\mathbf{H}_i^{(3)} : \widehat{\mathbf{a}}_0^{(3)}}{2 c_{\mathrm{s}}^6}
  \right)
\end{equation}
using Hermite coefficients \(\widehat{\mathbf{a}}_{0}^{(0)} = \rho(\bm{x},t)\) and \(\widehat{\mathbf{a}}_{0}^{(n)} = \mathbf{a}_{0}^{(n-1)} \widehat{\bm{u}}(\bm{x},t)\).

In the general case~\cite{KRAUSE20171}, we define the homogenized velocity \(\widehat{\bm{u}}\) as a convex combination of the fluid velocity moment \(\bm{u}\) and the solid velocity \(\bm{u}^{\mathrm{B}}\), given by  
\begin{equation}\label{eq:convexVelocity}
    \widehat{\bm{u}}(\bm{x},t) = (1 - d(\bm{x},t)) \bm{u}(\bm{x},t) + d(\bm{x},t) \bm{u}^{\mathrm{B}}(\bm{x},t), 
\end{equation}
where \(d\) is the so-called lattice porosity
\begin{equation}
    d(\bm{x},t) = 1 - \frac{\triangle x^2 \nu \tau_{\mathrm{mo}}}{K(\bm{x},t)}. 
    \label{eq:lattice_porosity}
\end{equation}
and \(\tau_{\mathrm{mo}}\) is the molecular relaxation time. While the the velocity of the solid geometry is set to zero for our urban flow cases, non-zero solid velocities are dynamically prescribed by the wall model in Section~\ref{sec:wm}.

Finally, the subgrid scale turbulence is accounted for by locally computing the effective relaxation time \(\tau_\mathrm{eff}(\bm{x},t)\) using the Smagorinsky BGK model
\begin{equation}\label{eq:tauEff}
    \tau_\mathrm{eff}(\bm{x},t) = \frac{\nu_\mathrm{eff}(\bm{x},t)}{c_{\mathrm{s}}^2} \frac{\triangle t}{\triangle x^2} + \frac{1}{2}.
\end{equation}

Connecting the HHRRLBM~\eqref{eq:hlbm} to the FBNSE~\eqref{eq:fbnse} target equation, we expect a second-order approximation in space for the fluid velocity moment \cite{SIMONIS23,simonis2023pde}.

\subsubsection{Turbulent Wall Model}\label{sec:wm}

It is computationally very expensive and basically infeasible to fully resolve the turbulent structures at the walls in large-scale urban flows.
An established approach to this problem is the use of wall modeling s.t. the velocity profiles and shear stresses in the boundary layer are modeled using \emph{universal turbulent velocity profiles}.
These functions characterize the streamwise mean velocity $u$ through the normalized variables
\begin{equation}\label{u+}
    u^+ = u \sqrt{\frac{\rho}{\tau_w}} = \frac{u}{u_{\tau_w}}
\end{equation}
and
\begin{equation}\label{y+}
    y^+ = \frac{y}{\nu}\sqrt{\frac{\tau_w}{\rho}} = y\frac{u_{\tau_w}}{\nu},
\end{equation}
where $\nu$ is the fluid kinemetic viscosity, $y$ the distance from the wall and $\tau_w$ the wall shear stress.
For the present case, we use the Spalding wall function which is valid for the range $y^+ \leq 1000$ and defined as
\begin{equation}\label{spaldingfunct}
    y^+ = u^+ + \frac{1}{E}\bigg[ e^{\kappa u^+} - 1 - \kappa u^+ - \frac{(\kappa u^+)^2}{2} - \frac{(\kappa u^+)^3}{6} \bigg],
\end{equation}
with a wall roughness parameter $E$.

The modeled velocity and shear stresses are incorporated into the HRRLBM as the velocity moment of the solid medium and by setting the hybridization factor \(\sigma\) to 0 (i.e. \(100\%\) FD rate of strain tensor) in the wall-modeled region.

\subsubsection{Turbulence Generation at the Inlet}

The air movements that interact with urban structures follow established \emph{atmospheric velocity profiles} that need to be prescribed at the inlet of the simulation.
One possibility of prescribing such perturbed profiles is via the \emph{Vortex Method} (VM) \cite{Hettel2025}.

Let \(u_\text{mean}, u_\text{prev} : \mathbb{R}^3 \to \mathbb{R}^3\) be the mean velocity profile and the inflow velocity of the previous timestep, \(A_\text{inlet} > 0\) the inlet area and \(d_\text{inlet} \in \mathbb{R}^3\) the normalized inflow direction.
Given the vortex size \(\sigma > \delta_x\) and turbulence intensity \(I > 0\) as parameters, \(n_\text{vortices} \in \mathbb{N}\) discrete seed points are placed randomly on the \emph{inflow plane} at positions \(p_i \in \mathbb{R}^3\) and associated with signs \(s_i \in \{-1,1\}\).
Based on this, per-vortex circulations \(\Upgamma_i\) are computed from turbulent kinetic energies \(k_i\):
\begin{align}
 k_i :&= \frac{3}{2} (|u_\text{mean}(p_i)|I)^2, \\
 \Upgamma_i :&= 4 s_i \sqrt{\frac{\pi}{6\ln{3}-9\ln{2}}\frac{k_i A_\text{inlet}}{n_\text{vortices}}}.
\end{align}
Together, this allows for recovering the pertubed velocity field
\begin{align}
   u_\text{vortex}(x) :&= \frac{1}{2\pi} \sum_{i=1}^{n_\text{vortices}} \Upgamma_i \frac{((p_i - x)\times d_\text{inlet})\left(1-\exp\left(\frac{|x-p_i|^2}{2\sigma^2}\right)\right)}{|x-p_i|^2}
\end{align}
which results in the final inflow velocity field with streamwise fluctuation using the Langevin equation
\begin{align}
 u(x) := u_\text{mean}(x) + u_\text{vortex}(x) -\frac{u_\text{vortex}(x) \times (\nabla u_\text{prev})(x)}{|(\nabla u_\text{prev})(x)|} d_\text{inlet}.
\end{align}
The pertubed inflow velocity field is computed for each discrete LB timestep and applied to the lattice using a regularized local velocity boundary condition. 

\subsection{Implementation in OpenLB}

At its core, OpenLB is a framework for efficiently and in parallel applying operators to distributed field-structured data on regular lattices.
Any such operator can transparently be executed both on CPU and GPU targets, transferring this ability to LB schemes implemented within this framework~\cite{kummerlaender25}.

\begin{figure}
\begin{subfigure}{0.23\textwidth}
\centering
\includegraphics[height=2.5cm]{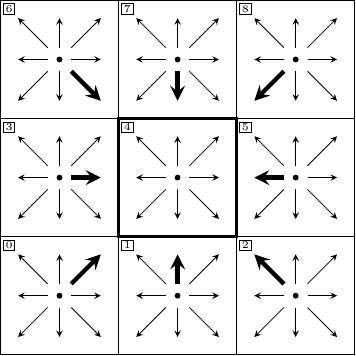}
\caption{Pre-stream lattice}
\end{subfigure}
\begin{subfigure}{0.23\textwidth}
\hspace*{-.5cm}
\includegraphics[height=2.5cm]{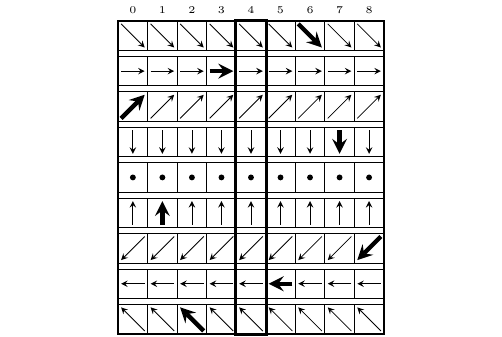}
\caption{Pre-stream data}
\end{subfigure}
\begin{subfigure}{0.22\textwidth}
\centering
\includegraphics[height=2.5cm]{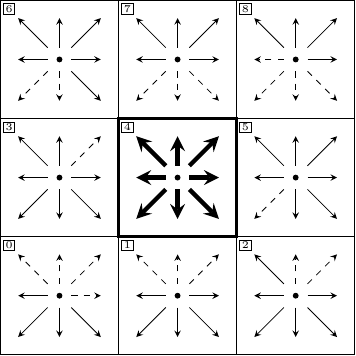}
\caption{Post-stream lattice}
\end{subfigure}
\begin{subfigure}{0.3\textwidth}
\includegraphics[height=2.5cm]{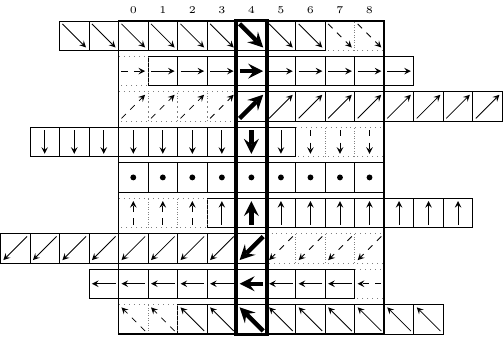}
\caption{Post-stream data}
\end{subfigure}
\begin{enumerate}[noitemsep,label=(\alph*)]
\item Bold populations are to be streamed into the center cell.
\item Bold rectangle encloses center cell populations.
\item Dotted arrows are values that rotated outside boundaries.
\item Bold arrows moved into the center rectangle by rotation.
\end{enumerate}
\caption{Implicit propagation without data-transfer using Periodic Shift (PS) \cite{kummerlanderImplicitPropagationDirectly2023}}
\label{fig:ps}
\end{figure}

A special case is the application of the LBM propagation step, the implementation of which is a critical component in the resulting performance~\cite{kummerlanderImplicitPropagationDirectly2023}. OpenLB utilizes an \emph{implicit} pattern that enables the propagation of populations between neighboring cells without copying but changing the view of the data. As depicted in Figure~\ref{fig:ps} this is achieved by storing the data in a vectorization friendly \emph{Structure of Arrays} (SoA) layout and rotating the array views within cyclic buffers.

\begin{figure}
    \includegraphics[width=\linewidth]{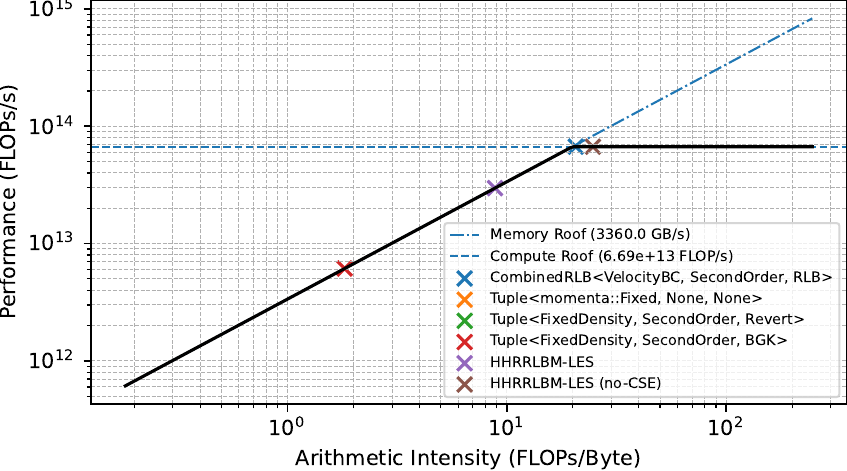}
    CSE optimization is critical for reducing the arithmetic intensity of HHRRLBM-LES, rendering the collision step bandwidth-limited in the ideal case. Memory and compute roofs given by theoretical maximums provided in the data sheets.
    \caption{Roofline analysis of local cell models for NVIDIA H100-94}
    \label{fig:roofline}
\end{figure}

The HHRRLBM~\eqref{eq:hlbm} collision step detailed in Section~\ref{sec:hhrrlbm} is implemented using OpenLB's dynamics tuple system, a kind of \emph{domain specific language} (DSL) for local LB cell models~\cite{kummerlaender25}.
Listing~\ref{lst:hhrrlbm} shows how the model is constructed as a tuple of moment system, equilibrium and collision operator.
\begin{listing}[h]
\begin{lstlisting}[language=myc++]
using HHRRLBM = dynamics::Tuple<
  T, descriptors::D3Q19<>, $\Comment{Value type and lattice stencil}$
  typename momenta::Tuple< $\Comment{Macroscopic moment system}$
    momenta::BulkDensity,
    momenta::MovingPorousMomentumCombination<momenta::BulkMomentum>,
    momenta::BulkStress,
    momenta::DefineToNEq
  >,
  equilibria::ThirdOrder, $\Comment{Equilibrium distribution}$
  collision::ParameterFromCell<$\Comment{Modified collision operator}$
    collision::LES::SMAGORINSKY,
    collision::SmagorinskyEffectiveOmega<collision::ThirdOrderRLB>>
>;
\end{lstlisting}
\caption{Dynamics tuple formulation of the HHRRLBM scheme}
\label{lst:hhrrlbm}
\end{listing}
Due to the abstract implementation of all LB models against the \emph{concept of a cell} they are not only amenable to platform-transparent execution but also automatic code optimization using common subexpression elimination (CSE).
For the present HHRRLBM model, OpenLB reports a per-collision memory bandwidth of 188 bytes and a floating point complexity of 1661 FLOPs (cf. 4644 FLOPs in the unoptimized case, a \(\sim 2.8\) fold reduction).
A theoretical roofline analysis for the involved dynamics utilizing OpenLB's introspection data is shown in Figure~\ref{fig:roofline}.

\section{Evaluation}\label{sec:evaluation}

\subsection{Parallel Efficiency}

We first investigate OpenLB's weak- and strong-scaling efficiency for the established lid-driven cavity benchmark case on all partitions of the HoreKa supercomputer.
OpenLB provides transparent execution for LB models across CPUs and NVIDIA GPUs, utilizing hybrid MPI+OpenMP+SIMD execution of CPU targets and MPI+CUDA on GPUs. The base arithmetic type is set to IEEE single-precision for all samples.

\begin{figure}
\medskip
\begin{subfigure}{\textwidth}
    \includegraphics[width=0.9\textwidth]{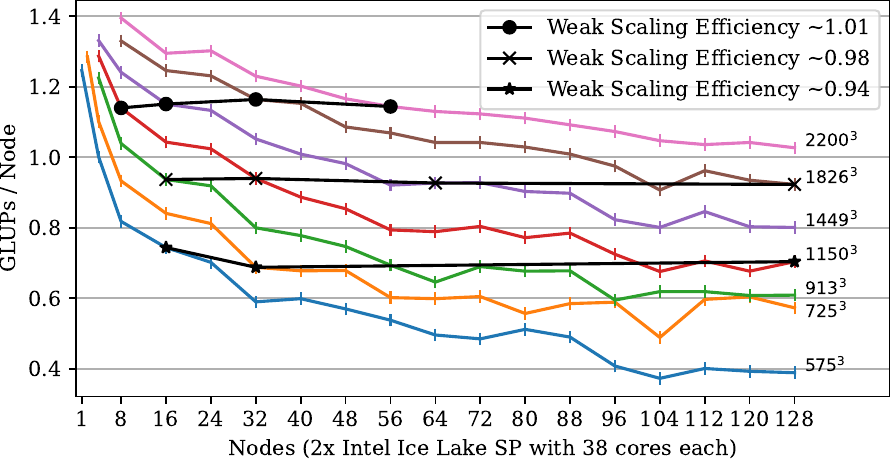}
    \caption{HoreKa Blue using MPI+OpenMP+AVX512}
    \label{fig:horekacpu}
\end{subfigure}

\medskip
\begin{subfigure}{\textwidth}
    \includegraphics[width=0.9\textwidth]{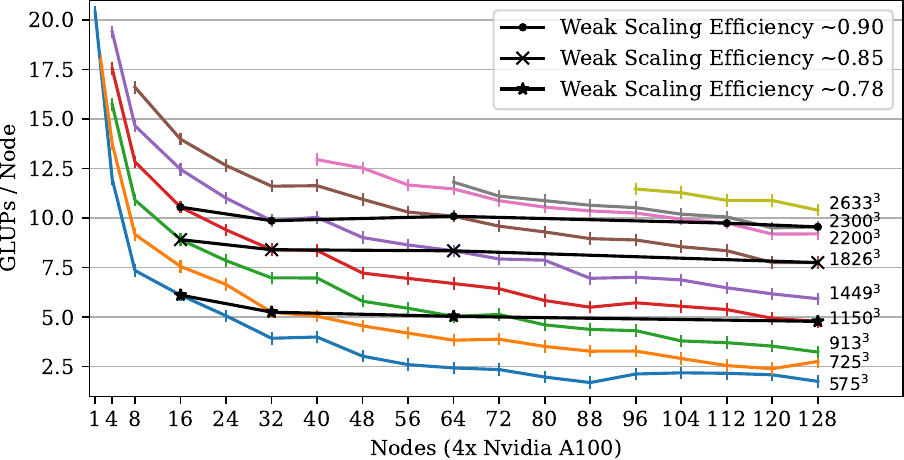}
    \caption{HoreKa Green using MPI+CUDA}
    \label{fig:horekagpu}
\end{subfigure}

\medskip
\begin{subfigure}{\textwidth}
    \includegraphics[width=0.9\textwidth]{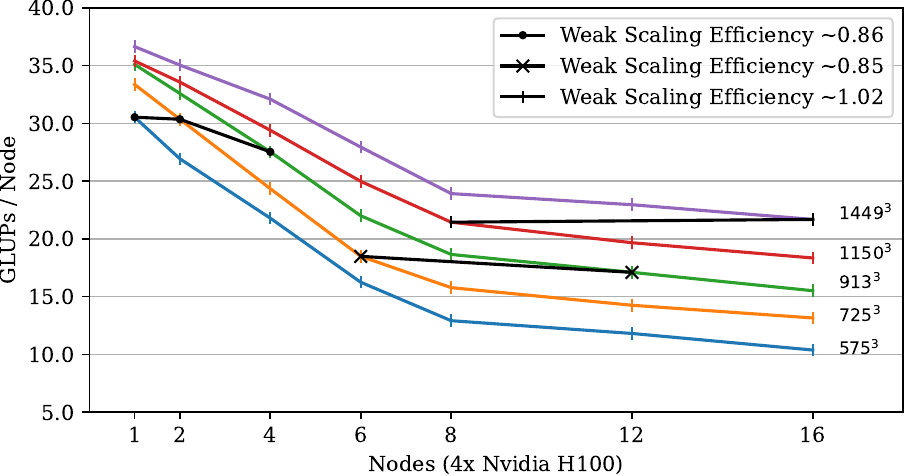}
    \caption{HoreKa Teal using MPI+CUDA}
    \label{fig:horekateal}
\end{subfigure}
\caption{Parallel efficiency of OpenLB on HoreKa}
\end{figure}

Figures~\ref{fig:horekacpu} and \ref{fig:horekagpu} are expanded versions of the previously published scalability study~\cite{KUMEERLAENDER22} on up to 128 nodes of HoreKa Blue resp. Green while Figure~\ref{fig:horekateal} details performance on up to 16 HoreKa Teal nodes (maximum number allocateable in \texttt{accelerated-h100}).
Problem sizes range between 190 million and 18 billion cells and strong efficiencies are listed additionally in Table~\ref{tab:strong_scaling_efficiencies}.

In all plots, the absolute and per-node performance is measured in \emph{billions of cell updates per second} (GLUPs). This is the established reference quantity for LBM performance measurements and it is directly related to the \emph{time-to-solution} by
\begin{equation}
    t_\text{solution}(N) = \frac{\Sigma_\text{cells}(N)\Sigma_\text{timesteps}(N)}{1e9 p(N)}
\end{equation}
for number of cells \(\Sigma_\text{cells}\), number of timesteps \(\Sigma_\text{timesteps}\) and performance measurement in GLUPs \(p\) for fixed resolution \(N\).

\begin{table}[h]
\centering
\begin{tabular}{l l l l l l l}
\toprule
Size & \multicolumn{2}{l}{HoreKa Blue} & \multicolumn{2}{l}{HoreKa Green} & \multicolumn{2}{l}{HoreKa Teal} \\
 & [32→128] & [64→128] & [32→128] & [64→128] & [4→16] & [8→16]  \\
\midrule
$575^3$  & 0.66 & 0.78 & 0.45 & 0.72 & 0.48 & 0.80 \\
$725^3$  & 0.83 & 0.96 & 0.53 & 0.72 & 0.54 & 0.83 \\
$913^3$  & 0.76 & 0.94 & 0.46 & 0.64 & 0.56 & 0.83 \\
$1150^3$ & 0.75 & 0.89 & 0.57 & 0.71 & 0.62 & 0.86 \\
$1449^3$ & 0.76 & 0.86 & 0.60 & 0.71 & 0.67 & 0.91 \\
$1826^3$ & 0.79 & 0.89 & 0.67 & 0.77 & — & — \\
$2200^3$ & 0.84 & 0.91 & —    & 0.80 & — & — \\
$2300^3$ & —    & —    & —    & 0.81 & — & — \\
\bottomrule
\end{tabular}
\caption{Strong scaling efficiencies on HoreKa for different problem sizes}
\label{tab:strong_scaling_efficiencies}
\end{table}

Summarizing the use of HoreKa for all applications within the project and not just the exemplary urban flow studies, we observe a continuing switch towards using a low number of accelerated nodes instead of many CPU-only nodes. Even a single HoreKa GPU node is sufficient for simulations up to \(\sim 1e9\) cells which is already beyond the needs of many practical applications.

\clearpage
\subsection{Urban Flow Reference Case}

For a full description and evaluation of the urban digital twin implemented using the HHRRLBM approach detailed in Section~\ref{sec:method} we refer to \cite{teutscher2025}.
Here, we summarize the validation against experimental and simulation data~\cite{GROMKE08}.

\begin{figure}
\centering
\includegraphics[width=0.75\textwidth]{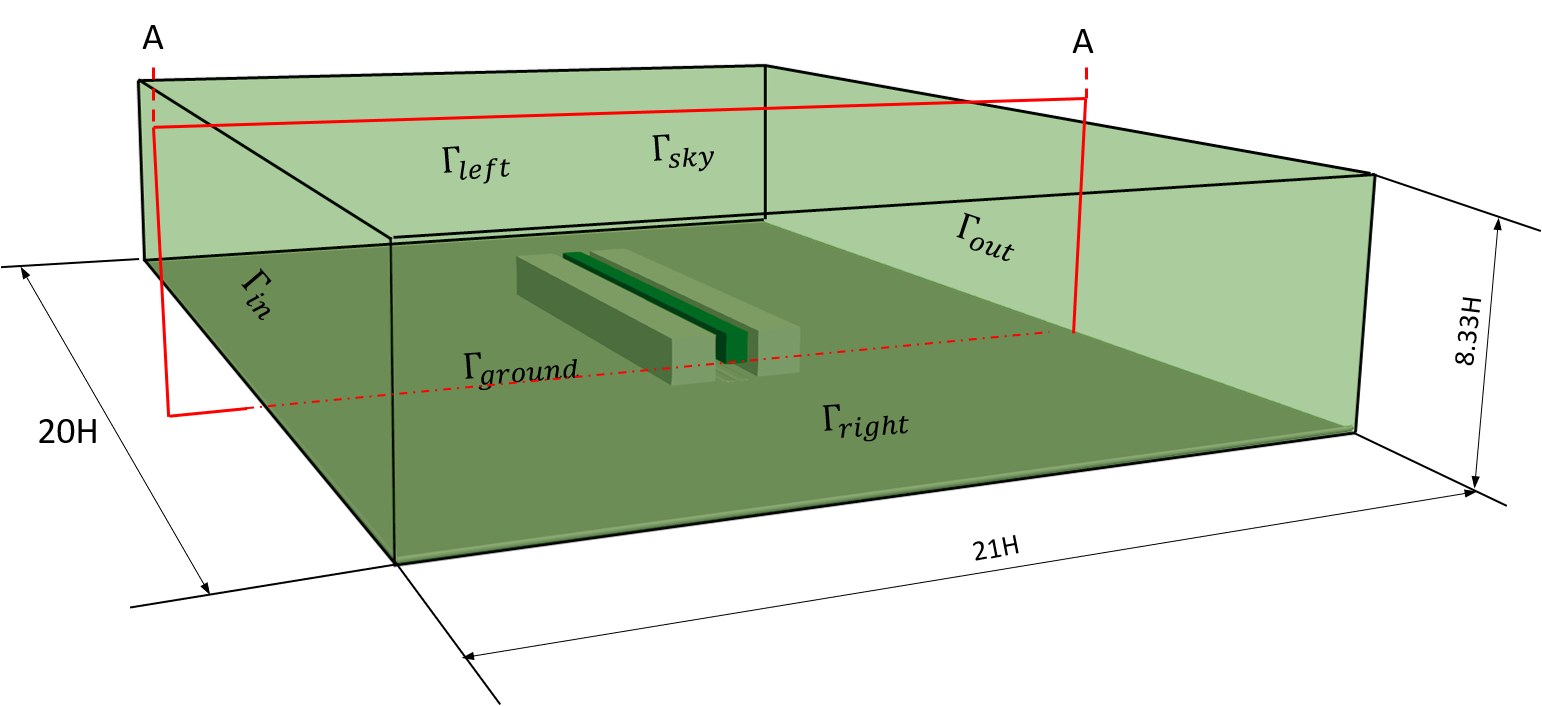}

\medskip
\begin{tabular}{l l l l l l l l}
\toprule
$Re \,[-]$ & $\triangle x \,[\text{m}]$ & $\triangle t \,[\text{s}]$ & $U_{\mathrm{H}} \,[\text{m}/\text{s}]$ & $U_e \,[\text{m}/\text{s}]$ &$T_{\mathrm{tot}} \,[\text{s}]$ & $H \,[\text{m}]$ & \(\Sigma_\text{cells} [-]\) \\
\midrule
$37000$ & $2 \times 10^{-2}$ &$1.8\times 10^{-5}$ & $4.65$ & $0.2054$&$25$&$0.12$ &$439\times10^6$\\
\bottomrule
\end{tabular}
\caption{Simulation setup for reproducing the experimental reference case~\cite{GROMKE08,teutscher2025}}
\label{fig:geometrical_setup}
\end{figure}

The geometric setup depicted in Figure~\ref{fig:geometrical_setup} represents two rows of buildings forming a \emph{street canyon}.
In OpenLB, the atmospheric inlet profile~\cite{PASQUIER23} is modeled using the VM, the outlet using a fixed local pressure condition with fringe zone and the outer walls using full slip boundaries.  

\begin{figure}
    \begin{subfigure}{0.3\textwidth}
        \begin{tikzpicture}
            \begin{axis}[
                axis on top,
                xlabel={$x/H$},
                ylabel={$z/H$},
                ylabel style={yshift=-10pt},
                width=1.6\textwidth, 
                xmin=-0.5, xmax=0.5,
                ymin=0, ymax=1.3,
                minor grid style={dotted},
                xtick={-0.4,-0.2,0,0.2,0.4},
                ytick={0,0.2,0.4,0.6,0.8,1,1.2},
                minor xtick={-0.3,-0.1,0.1,0.3},
                minor ytick={0.1,0.3,0.5,0.7,0.9,1.1},
                axis equal image,
            ]
                \addplot graphics[xmin=-0.5, xmax=0.5, ymin=0, ymax=1.3] {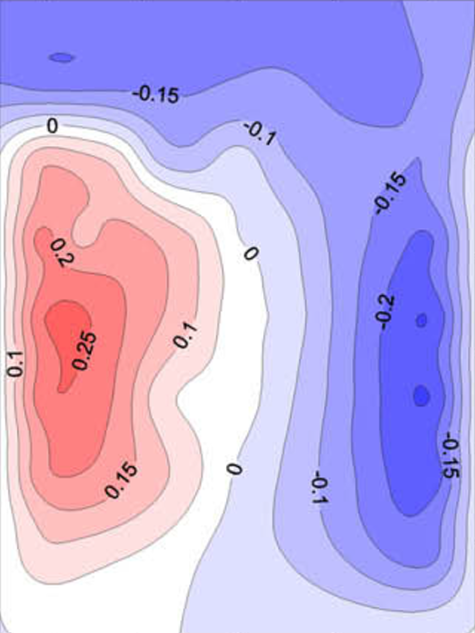};
            \end{axis}
        \end{tikzpicture}
        \caption{Experiment~\cite{GROMKE08}}
    \end{subfigure}
    \begin{subfigure}{0.3\textwidth}
            \begin{tikzpicture}
                \begin{axis}[
                    axis on top,
                    xlabel={$x/H$},
                    ylabel={$z/H$},
                    ylabel style={yshift=-10pt},
                    width=1.6\textwidth,
                    xmin=-0.5, xmax=0.5,
                    ymin=0, ymax=1.3,
                    minor grid style={dotted},
                    xtick={-0.4,-0.2,0,0.2,0.4},
                    ytick={0,0.2,0.4,0.6,0.8,1,1.2},
                    minor xtick={-0.3,-0.1,0.1,0.3},
                    minor ytick={0.1,0.3,0.5,0.7,0.9,1.1},
                    axis equal image,
                ]
                    \addplot graphics[xmin=-0.5, xmax=0.5, ymin=0, ymax=1.3] {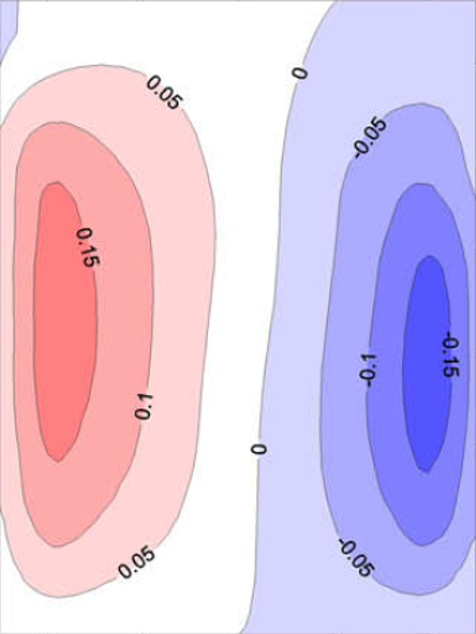};
                \end{axis}
            \end{tikzpicture}
            \caption{FVM~\cite{GROMKE08}}
        \end{subfigure}
        \begin{subfigure}{0.3\textwidth}
            \begin{tikzpicture}
                \begin{axis}[
                    axis on top,
                    xlabel={$x/H$},
                    ylabel={$z/H$},
                    ylabel style={yshift=-10pt},
                    width=1.6\textwidth,
                    xmin=-0.5, xmax=0.5,
                    ymin=0, ymax=1.3,
                    minor grid style={dotted},
                    xtick={-0.4,-0.2,0,0.2,0.4},
                    ytick={0,0.2,0.4,0.6,0.8,1,1.2},
                    minor xtick={-0.3,-0.1,0.1,0.3},
                    minor ytick={0.1,0.3,0.5,0.7,0.9,1.1},
                    axis equal image,
                ]
                    \addplot graphics[xmin=-0.5, xmax=0.5, ymin=0, ymax=1.3] {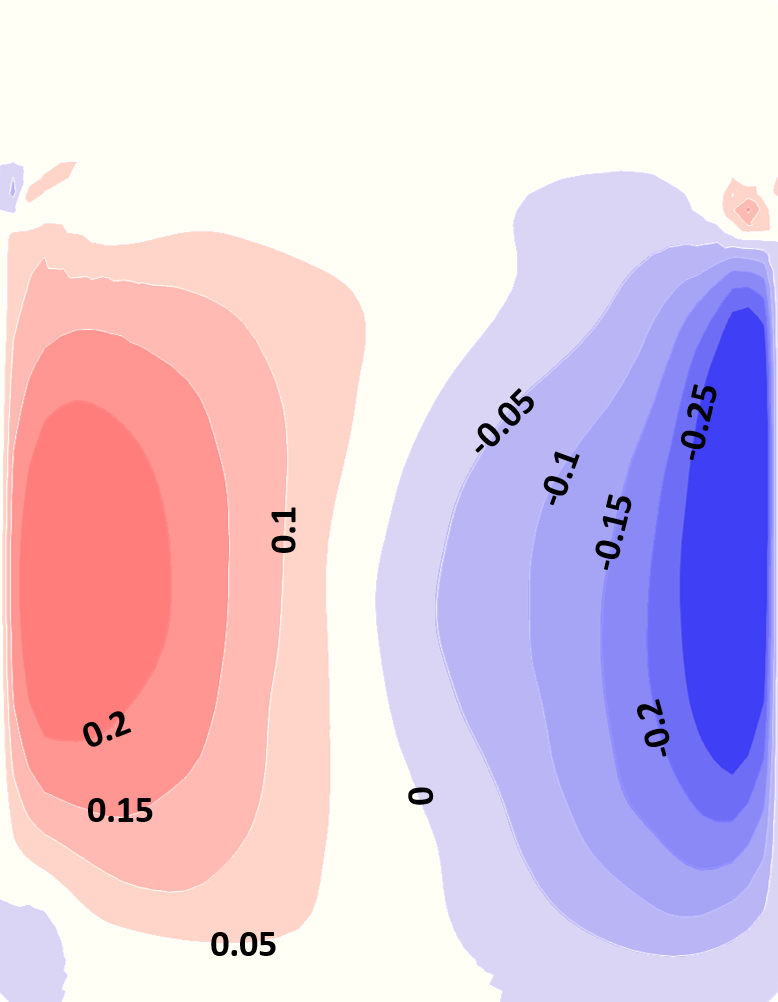};
                \end{axis}
            \end{tikzpicture}
            \caption{LBM~\cite{teutscher2025}}
        \end{subfigure}

Normalized vertical velocity magnitude for the canyon reference case along plane cut A-A (cf. Figure~\ref{fig:geometrical_setup}). FVM results obtained using FLUENT~\cite{GROMKE08} and LBM results using OpenLB~\cite{teutscher2025}. Both simulations used turbulent wall modeling. LBM results obtained using \(438e6\) cells on HoreKa Teal. 
    \caption{Comparison of experimental and simulation results for the canyon case~\cite{GROMKE08}}
    \label{fig:valid}
\end{figure}

\begin{figure}
    \includegraphics[width=0.9\textwidth]{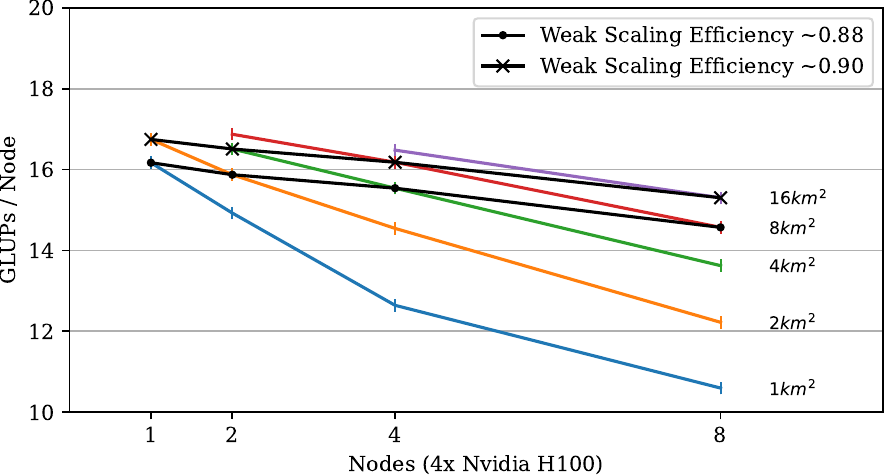}
    
    Single-precision simulations of the FBNSE-only numerical wind channel case including VM turbulent inlet condition and wall modeling. City sections between 1 and \(16 \text{km}^2\) are resolved with a voxel size of \(1 \text{m}\) yielding problem sizes between \(0.5 \times 10^9\) and \(8 \times 10^9\) cells.
    \caption{Parallel efficiency of the numerical wind channel on HoreKa}
    \label{fig:cityscale}
\end{figure}

Based on the validation of the basic simulation setup summarized in Figure~\ref{fig:valid}, a full-scale version of the numerical wind channel was used for building a digital twin~\cite{teutscher2025} incorporating real-world pollution measurements.
Figure~\ref{fig:cityscale} show a combined weak- and strong-scaling study of the numerical wind channel in OpenLB for city sections of up to \(16 \text{km}^2\). Despite the lower absolute performance explained by the significantly more complex simulation setup, parallel efficiency on HoreKa Teal compares well to the lid-driven cavity benchmark results in Figure~\ref{fig:horekateal}.

\section{Conclusion}

HHRRLBM-LES was described as a parallelization-friendly discretization for the filtered Brinkmann-Navier-Stokes equations. Its efficient implementation for platform-transparent execution on both SIMD CPUs (using MPI+OpenMP+AVX512) an GPUs (using MPI+CUDA) in OpenLB was detailed. Automatic code optimization was discussed and the performance of local cell models was studied using a roofline analysis.
Extensive strong and weak scaling evaluations with efficiencies up to \(1.02\) for problem sizes up to 18 billion cells were provided. Transferability of parallel efficiency from ideal benchmark cases to complex wall-modeled turbulent flow cases was demonstrated.

In summary, this showcases how OpenLB can deliver on the promise of fast large-scale turbulent flow simulations on state-of-the-art supercomputers.

\section*{Resource Usage}

As the budgets are not separated between the current and past \texttt{CPE} project in the \texttt{kit\_project\_usage} reporting command we can only provide total used resources:
\begin{lstlisting}[language=introspection]
Project		Resource	Granted		Used		Percent
hk-project-cpe	cpu		81932606	72868631.5	88.93%
hk-project-cpe	gres/gpu	719495		848282		117.89%
\end{lstlisting}

\bibliographystyle{elsarticle-num} 
\bibliography{references.bib}

\end{document}